\documentclass[11pt,twoside]{article}
\usepackage{asp2010}
\resetcounters

\begin{document}

\title{DECal: A Spectrophotometric Calibration System For DECam}
\author{J. L. Marshall, Jean-Philippe Rheault, D. L. DePoy, 
Travis Prochaska, Richard Allen, Tyler W. Behm, Emily C. Martin, Brannon Veal, Steven Villanueva, Jr., 
Patrick Williams, and Jason Wise
\affil{Department of Physics and Astronomy, Texas A\&M University, College Station, TX 77843-4242}}

\begin{abstract}
DECal is a new calibration system for 
the CTIO 4 m Blanco telescope.  It is currently being installed 
as part of the Dark Energy Survey and will provide both broadband flat fields 
and narrowband ($\sim$1 nm bandwidth) spectrophotometric calibration for the new Dark Energy Camera (DECam). 
Both of these systems share a new Lambertian flat field screen.  
The broadband flat field system uses LEDs to illuminate each photometric filter. 
The spectrophotometric calibration system consists 
of a monochromator-based tunable light source that is projected onto the
flat field screen using a custom line-to-spot fiber bundle 
and an engineered diffuser. Several calibrated photodiodes positioned 
along the beam monitor the telescope throughput as a function of wavelength. 
This system will measure the wavelength-dependent instrumental response function 
of the total telescope+instrument system in the range 300 $<\lambda<$ 1100nm. The 
spectrophotometric calibration will be performed regularly (roughly once per month) to determine the spectral response of the 
DECam system and to monitor changes in instrumental 
throughput during the five year Dark Energy Survey project. 
\end{abstract}

\section{Introduction}
The Dark Energy Survey (DES) is a next generation
deep, wide, multi-band imaging survey that will map 5000 square degrees 
of sky in 525 nights over five years beginning in late 2012.  DES relies on the new 
570-Megapixel, 2.2 degree field-of-view 
Dark Energy Camera, known as DECam \citep{decam}, now installed at the prime focus
of the 4 m Blanco telescope at the Cerro Tololo Inter-American Observatory (CTIO).  

Modern imaging surveys require highly accurate and precise photometry to meet stringent requirements placed on the data 
by the science goals.
The requirement for accurate and precision photometry in turn requires accurate and precise calibration of the data products.  
For example, DES has a goal of achieving photometry accurate to 1\% (0.01 mag)
over the entire survey region.  
In order to make measurements with this level of accuracy, care must be taken to 
calibrate the data appropriately and to remove any spurious effects due to 
weather, the earth's atmosphere, noise and other features of 
the detector system, and changes in instrumental throughput or response.
These calibrations must be made in addition to standard astronomical photometric corrections 
such as photometric zeropoint corrections and airmass and color terms.

DECal is a complete calibration system for DECam that will 
allow for the correction of two of these sources of error.  DECal will
provide daily flat fields for DECam to correct the pixel-to-pixel variations across the 
CCD detectors; it will also measure the relative throughput
of the complete telescope+instrument system as a function of wavelength. 
This spectrophotometric calibration system will allow for the monitoring of the 
instrumental throughput (e.g., to determine whether optical coatings evolve with time); 
it will also provide accurate knowledge of the filter transmission functions
that can be used to calculate precise Supernova k-corrections and improved photometric redshift measurements. 
We plan to use DECal to monitor the throughput of the telescope at regular
intervals (i.e., $\sim$once per month) during the five-year survey to monitor the
instrumental performance.
DECal will be installed permanently at the Blanco telescope and will be available for general use. 

Similar spectrophotometric calibration systems have been tested and
installed on the CTIO 4 m Blanco telescope \citep{stubbs2007} 
and the PanSTARRS telescope \citep{panstarrs}.
Both of these systems use a tunable laser as a light source. The DECal system uses a 
monochromator-based light source in place of a tunable laser, which 
generally requires less maintenance and other personnel attention than a tunable 
laser, and is well suited for routine use over an extended period of time. 
The DECal system also has the advantage of its extended and continuous 
wavelength coverage.  A system such as DECal can operate from the far UV into the infrared (250 $<\lambda<$ 2400 nm) 
with relatively minor modifications.  The disadvantage 
of a monochromator-based light source is that it cannot produce as much light
as a tunable laser, so integration times must be longer and the dome must be kept very dark.  

We have successfully deployed a prototype of the DECal spectrophotometric calibration system at the Swope 
1 m and du Pont 2.5 m telescopes at Las Campanas Observatory in Chile. 
We measured the throughput of the $u$, $g$, $r$, $i$, $B$, $V$, $Y$, $J$, $H$, and $K_s$ filters used in the WIRC and RetroCam instruments
during the Carnegie Supernova Project with an accuracy of 
1\% \citep{stritz}.
A forthcoming paper on the completed DECal system will present the results of the calibration 
of the DECam instrument using the final DECal system once both DECam and DECal are installed at the CTIO 4m telescope.  

\begin{figure}
\plotone{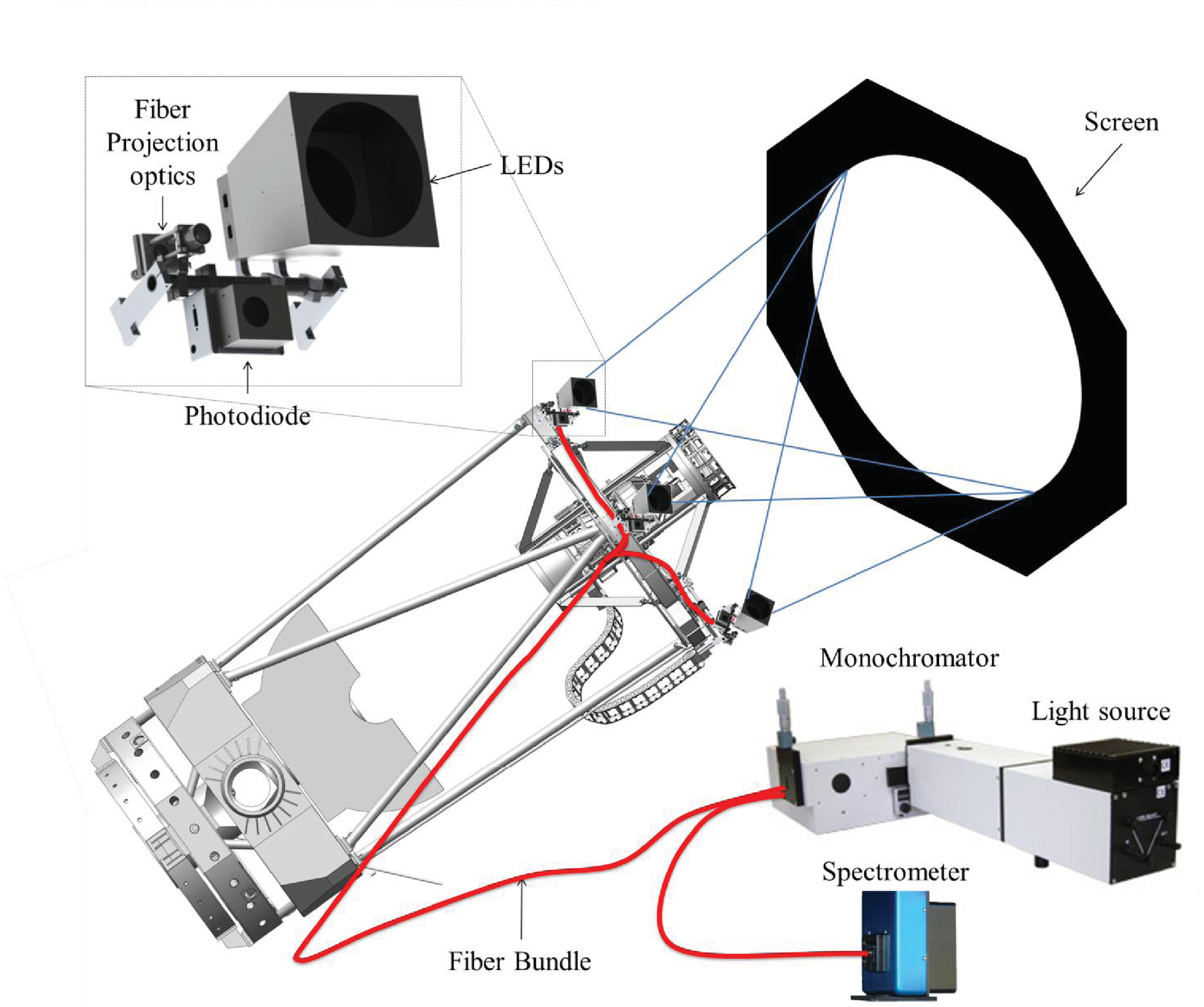}
\caption{Schematic drawing of the DECal system.
\label{fig:overview}
}
\end{figure}

The complete DECal system consists of a new 
flat field screen, a daily dome flat field illumination system, and 
the spectrophotometric calibration system.  These systems are described in detail below.
A schematic of the entire calibration system is shown in Figure \ref{fig:overview}.
More details on the final system may be found in a recent paper by \citet{decal2012}.

\section{Flat Field Screen}

The flat field screen is an important part of any dome-based calibration
system and should be carefully considered when designing the calibration scheme.  
An ideal flatfield screen would have 100\% reflectivity at all wavelengths 
of interest and would reflect light only within the acceptance angle of the 
telescope optics. Since the latter condition is almost impossible to satisfy, 
we have chosen to provide uniform illumination of the top end of the telescope
by producing an illumination pattern with a Lambertian profile.  
A Lambertian surface reflects incident light such that  
the surface luminance of the screen as seen by the telescope is isotropic.   A flat field screen with Lambertian scattering
properties ensures that all points on the screen illuminate the focal plane
with the same angular profile regardless of where the illumination source is 
placed. If the telescope is well baffled, illumination of this type should be 
equivalent to illuminating the telescope with incident light at only the acceptance angle of the telescope.

The DECal flat field screen is made of a 2x4 grid of 4 foot by 8 foot 
lightweight aluminum honeycomb panels coated with 
a white, highly reflective, almost perfectly Lambertian coating.   
These panels are mounted on an extruded aluminum structure using screws that have 
been coated with the same coating as the screen.  
The white circular portion of the screen is slightly oversized (4.64 m) and is surrounded by a black ring made of 
sheet metal painted with flat black spray paint by Maaco (note that we do not recommend this 
vendor for black coatings).  Our studies show that it is important to 
include this outer ring on the flat field screen to minimize stray light reflected into the beam by the 
shiny internal surface of the dome.  We have also determined that any stray light 
reflected by the seams between the two panels or the heads of the screws is minimal 
and has almost no effect ($<<$1\%) on the precision of photometric measurements that 
have been flattened with this system; these results will be presented in a forthcoming paper describing the 
deployed DECal system.

Several candidate flat field screen coatings were tested for absolute reflectivity as well as reflectivity 
as a function of incidence angle (Lambertian-ness).
More details of this study are provided by \citet{decal2010}.  
Out of several candidate screen coatings we finally selected the nearly ideal 
``Duraflect'' coating provided by Labsphere, which provides high reflectivity 
(roughly 95\% from 350 to 1200nm and greater than 85\%  for 300 $<\lambda<$ 2200nm).
This coating also boasts a nearly perfectly Lambertian scattering surface.  It is relatively durable and can be 
(gently) cleaned, an important quality for the dome environment.

\section{Dome Flat Field Illumination}

\begin{figure}
\plotone{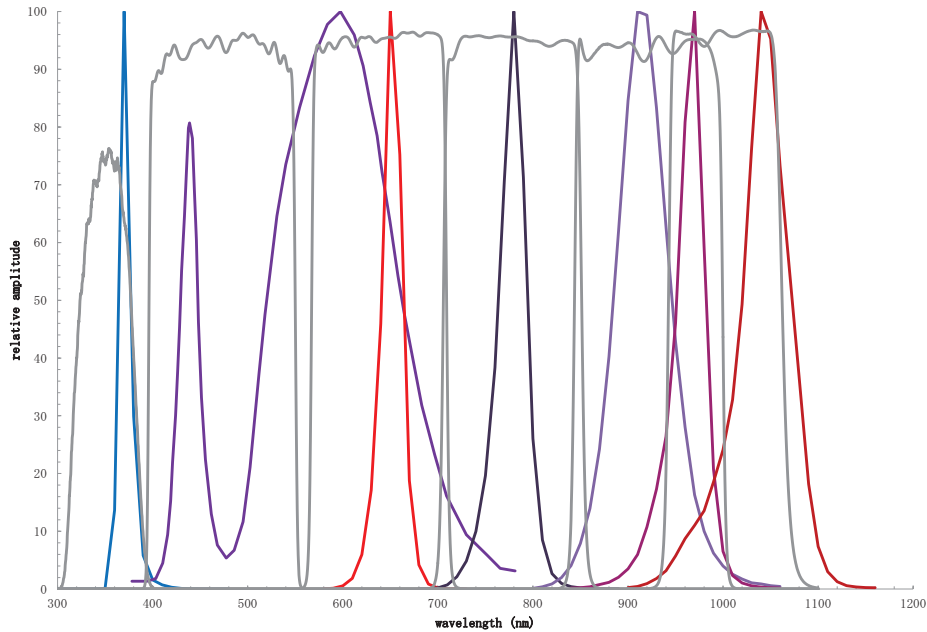}
\caption{Spectra of selected DECal LEDs overplotted with the DES $grizy$ (and $u$) filter response curves. The normalization is arbitrary.
\label{fig:led}
}
\end{figure}

The DECal daily flat field calibration system provides for daily flat fields to be taken at 
the telescope.  It uses high-power light-emitting diodes (LEDs) as the illumination source.  
We have selected one LED to illuminate each filter bandpass used in the DES survey,
usually centered on the filter bandpass.   We have also provided LEDs at a wavelength that can 
illuminate a planned u-band filter, which will be added to DECam by CTIO in the near future. Figure \ref{fig:led}
shows the spectra of the selected LEDs along with the DES filter bandpasses.

The seven LEDs selected to illuminate the DECam filter bandpasses are manufactured by Roithner Lasertechnik and Luxeon Star.  
The model numbers of the LEDs are given in Table \ref{table:leds}.
They are positioned at four locations around the top of the telescope ring and provide adequately flat illumination 
of the flat field screen.  
The flat field screen does not need to be perfectly uniformly illuminated to produce 
a reasonably uniform illumination of the focal plane; however, large scale gradients 
should be avoided to minimize the need to remove such trends from the data.  
The calibration system 
can be tuned so that the relative power in each bandpass is about the same, i.e., DECam 
exposure times will be the same for each filter.

\begin{table}[!ht]
\label{table:leds}
\caption{DECal LEDs}
\smallskip
\begin{center}
{\small
\begin{tabular}{lll}
\tableline
\noalign{\smallskip}
Manufacturer & Model Number & Central Wavelength\\
\noalign{\smallskip}
\tableline
\noalign{\smallskip}
Roithner & H2A1-H365-S & 365 nm\\
Roithner & H2A1-H650 & 650 nm\\
Roithner & H2A1-H780 & 780 nm\\
Roithner & H2A1-H905 & 905 nm\\
Roithner & H2A1-H970 & 970 nm\\
Roithner & H2A1-H1030 & 1030 nm\\
Luxeon Star & Warm white (3100k) Rebel LED & broad\\
\noalign{\smallskip}
\tableline
\end{tabular}
}
\end{center}
\end{table}

Testing of similar LED-based dome flat field systems \citep{ff2005} shows that it 
is not necessary to illuminate the entire filter bandpass to provide adequate 
flat fields.  The exception to this, of course, is in the case that the detector
quantum efficiency changes rapidly with wavelength.  This is usually not 
the case for modern CCD detectors.

\section{Spectrophotometric Calibration System}

In addition to broadband flat fields, DECal also projects nearly monochromatic (1--10nm bandwidth)
flat field light onto the flat field screen to be used to measure the relative throughput of the instrument 
as a function of wavelength.  This spectrophotometric calibration is accomplished by 
imaging the monochromatic light incident on the flat field screen with DECam while at the same time
monitoring the amount of light on the screen with calibrated photodiodes placed around the top ring of the telescope.  
The signal received by the photodiodes is proportional to the light received by DECam; note however that we do not attempt to 
estimate the total amount of light gathered by the telescope, i.e., 
this is a relative calibration, not an absolute measure of instrumental throughput.
The DECam images are compared with the photodiode output to determine the 
relative sensitivity of the entire telescope+instrument optical system as 
a function of wavelength.  
This part of the DECal system will be used to spectrophotometrically calibrate 
the DES survey data, and to monitor the DECam instrument's sensitivity as a function of time.

We calculate that this system will have a peak output power of 2 mW resulting in approximately 
800 photons/second/pixel when imaged by DECam.  This flux level will require exposure times with DECam of 
approximately one minute to obtain adequate signal-to-noise ratio (S/N) measurements
of the DECam focal plane; given the relatively bright environment of the Blanco dome we anticipate that 
the DECal measurements will best be made during a cloudy night: to minimize scattered light from the Sun during the daytime and 
to avoid interfering with regular nighttime observations.

\subsection{Monochromator}
The monochromatic light is produced by a fully automated Czerny-Turner monochromator 
(manufactured by Horiba, model iHR-320). We adjust the 
bandwidth of the light by varying the input slit width. The output slit width is fixed to 
900 microns (the width of the linear fiber bundle that feeds light to the top of the telescope). 
The monochromator holds two light sources, multiple gratings on a turret, and all of the necessary 
order sorting filters to properly produce the monochromatic light at the appropriate wavelength. 
It is completely remotely controlled by a Labview interface developed in our lab and is highly flexible 
in producing light of the appropriate wavelength, bandwidth, and interval between requested wavelengths. 

\subsection{Fiber bundle}
We have designed and manufactured a custom 75 m long ``line-to-spot'' fiber bundle (assembled by Fibertech Optica) that 
uses a broad-spectrum fiber (Polymicro FBP-300660710). This fiber has excellent 
transmission both in the UV and the IR, in contrast with standard optical fibers that have either good 
transmission  in the infrared (due to low OH$^-$ content) or in the UV 
(high OH$^-$ content) but not at both ends of the spectrum simultaneously. 

The bundle consists of 87 fibers with 300 micron cores. At the input end, the fibers are 
arranged in 3 parallel lines of 29 fibers each.  The bundle brings the light from 
the calibration room located beneath the telescope main floor through the telescope cable wrap and
to the telescope top ring. At a 
length of 65 m, the fiber bundle splits into four branches, each 10 m long and containing 
21 fibers. These 21 fibers are placed in a compact circular arrangement and coupled to the
projection system. Each of the four branches sample the monochromator slit evenly to ensure 
that they all have the same intensity and spectral content.

A fifth branch bifurcates from the main bundle 1 m from the input slit. It contains the 
three central fibers and is fed to the monitor spectrometer (see below). To ensure that 
any absorption present in the fiber does not change the measured wavelength profile, 
this fifth branch is also 75 m long even though the spectrometer is located only 1 m from the monochromator. 

\subsection{Projection system}

To ensure a uniform 
illumination of the focal plane area, the output from the four fiber ends is collimated 
and then an engineered diffuser (manufactured by RPC Photonics)
is used to diffuse the light projected onto the flat field screen by the fibers. 
This type of diffuser can be designed 
to distribute incident collimated light in almost any pattern; in this case 
we chose a 20 degree angle cone diffuser that projects the light in a top hat shape with a 
diverging half-angle of 20 degrees, with more than 80\% of the 
light exiting the fiber bundle falling within this 40 degree cone of light. Engineered diffusers such as the one used in DECal are a far 
superior way to diffuse incident light as compared to a standard ground glass diffuser that would also produce an adequately
diffuse light pattern but would scatter much of the light away from the screen. 

\subsection{Spectrometer}

To monitor in real time the spectral content of the illumination source, 
a branch of three fibers bifurcates from the main fiber bundle and is fed to an echelle spectrometer 
(manufactured by Optomechanics Research, model SE-100). 
The spectrometer measures both the central wavelength 
and the full width at half maximum (FWHM) of the light. For each DECal exposure, a spectrum of the DECal light is obtained and automatically analyzed 
to measure the central wavelength and FWHM of the illumination light with a precision of 0.1 nm. 
The spectrometer calibration is verified with a Mercury calibration lamp.

\subsection{Reference photodiodes}

The goal of the spectrophotometric calibration system is to measure the relative throughput of the 
telescope+instrument optical system at each wavelength.  This is accomplished by imaging the 
monochromatic light incident on the flat field screen with the DECam imager while at the same time 
monitoring the light on the screen to remove any fluctuations in brightness.  
The monitoring of the screen is done with the reference photodiodes.  

The reference photodiodes are mounted 
in the same locations as the fiber projection units at four points around the top ring of the telescope, facing the
screen (see insert on Figure \ref{fig:overview}).   We use 10 mm diameter 
silicon photodiodes (manufactured by Hamamatsu, model S2281) to measure the light in the range 300 $<\lambda<$ 1100nm.
We acquire the signal from the photodiodes with a data acquisition unit and record the data with the Labview interface.

The photodiodes we use exhibit a slight change in sensitivity with temperature 
at the far red end of their sensitivity range. This effect is negligible below 1000nm but can reach a 
sensitivity change of up to 1\%/\deg C at 1100nm. We monitor the photodiode temperature in real time 
and correct for it in the data reduction process.

\section{Results}

We have used a prototype of the DECal spectrophotometric calibration system to measure the 
system throughput as a function of wavelength for the WIRC infrared imager on the 2.5 m du Pont telescope and Retrocam optical and infrared imager on the 1 m 
Swope telescope at Las Campanas Observatory, to improve the calibration of the 
Carnegie Supernova Project (CSP) data obtained on these telescopes.
A description of the DECal prototype used for these measurements is presented by \citet{decal2010} and the 
results of the successful calibration are applied to the optical CSP data and presented by \citet{stritz}.
In brief, the DECal prototype used at the du Pont and Swope telescopes included a new flat field screen for each telescope 
with the same properties but smaller in size than the flat field screen described above, a monochromator smaller but functionally the same 
as the final DECal monochromator, a custom line-to-spot fiber bundle using the same high throughput fibers used in DECal, 
and a monitor photodiode mounted behind the secondary mirror of each telescope.
Here we present the results of the optical and infrared measurements made using the DECal prototype.

\begin{figure}
\plotone{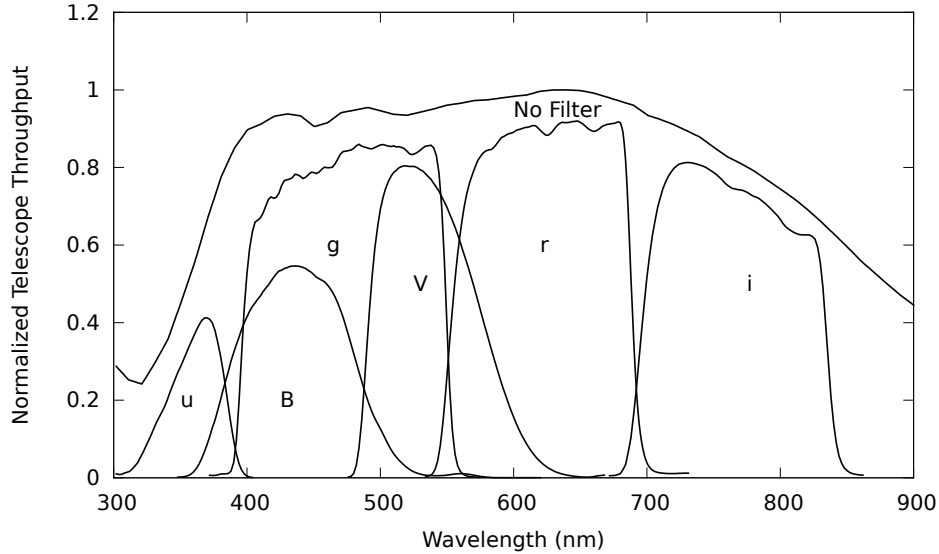}
\caption{DECal prototype scan of the optical filters used by the CSP survey in the Retrocam instrument on the Swope telescope.
\label{fig:results_optical}
}
\end{figure}

Figure \ref{fig:results_optical} shows the scans of 
the six optical filters used to make CSP photometric measurements on 
the Swope telescope, as well as the scan we made of the telescope+instrument system with no filter in place.   
The figures presented in this section give the relative throughput of the telescope, including losses by the primary and secondary mirrors, 
corrector plate, filter, dewar window and CCD quantum efficiency.  

The scan of the CSP Retrocam $u$-band filter was particularly 
useful in analyzing the CSP photometric data.  The manufacturer of the $u$-band filter did not provide 
a complete throughput curve with the filter.  An assumption (which turned out to be not entirely correct)
was made about the spectral response of the blue edge of the $u$-band filter in the initial data reduction of CSP data; 
we were able to measure the true spectral response of the $u$-band filter to thereby improve the CSP photometry with our measurements.

\begin{figure}
\plotone{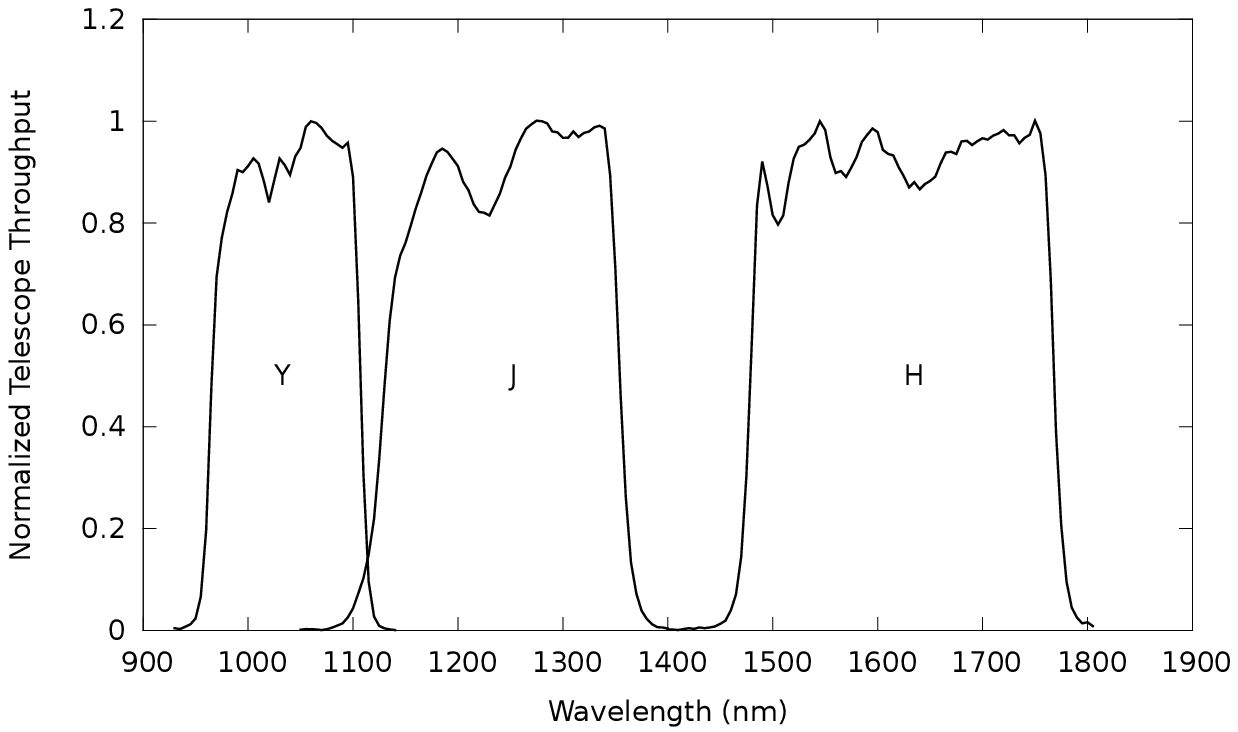}
\caption{DECal prototype scan of the infrared filters used by the CSP survey in the Retrocam instrument on the Swope telescope.
\label{fig:results_swopeir}
}
\end{figure}

\begin{figure}
\plotone{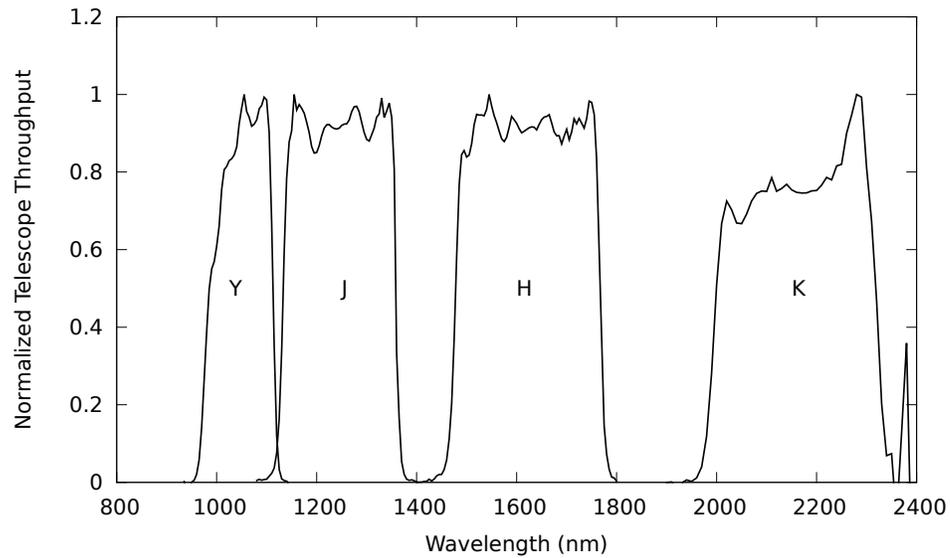}
\caption{DECal prototype scan of the infrared filters used by the CSP survey in the WIRC instrument on the du Pont telescope.
\label{fig:results_irdupont}
}
\end{figure}

Figures \ref{fig:results_swopeir} and  \ref{fig:results_irdupont} show scans of the infrared CSP 
bandpasses used on the Swope and du Pont telescopes, respectively.  
Each infrared filter scan has been normalized independently, i.e., the relative throughput from filter to filter is not 
reflected in these figures.
These data will also be used to calibrate the CSP infrared photometric measurements, to be published in a forthcoming 
CSP data release.
The infrared system scan measurements had slightly lower signal-to-noise (S/N) than the optical measurements; 
this is the source of the noise seen at the red end of the $K_s$-band data in Figure \ref{fig:results_irdupont}.
However, we note that with multiple measurements or increased signal from a brighter light source or even a higher throughput 
monochromator, higher S/N measurements of the infrared filters could be made. 
That is, a system such as DECal could be used to determine the spectral response of an imaging system well into the infrared
with only minimal modification.

\section{Summary}

DECal is a new calibration system that provides broadband and monochromatic flat fields for the DECam instrument 
on the CTIO Blanco  4 m telescope.  DECal will
be used to flatten optical DECam images and to monitor the transmission function of the instrument+telescope system 
and provide spectrophotometric calibration of the system.

We have fully tested the components of DECal and have deployed a prototype of the system on the Las Campanas Observatory 
Swope and du Pont telescopes.  This prototype has informed the design of the final 
DECal system and was used to successfully calibrate the Carnegie Supernova Project optical and infrared photometry.

The DECal and DECam systems are currently in the final stages of installation on the CTIO Blanco 4 m telescope and will
be commissioned in Fall 2012.  We will use the DECal broadband flat field system to produce daily flat fields 
for the DECam instrument, and the DECal spectrophotometric calibration system to produce monthly scans of the 
spectral response of the complete telescope+instrument system, through each DES filter and at all wavelengths.  
These data will be used to measure the 
spectral response of the instrument, which will enable the highly accurate and precise photometric measurements required by DES.  
The DECal system will also be used to monitor any changes in the spectral response of the instrument, 
including evolution of filter functions, degradation of optical coatings, and off-band optical leaks in the filters.

\acknowledgements Texas A\&M University thanks Charles R. '62 and Judith G. Munnerlyn, 
George P. '40 and Cynthia Woods Mitchell, and their families for support of astronomical 
instrumentation activities in the Department of Physics and Astronomy.

\bibliography{marshall_j}

\begin{thebibliography}{}
\expandafter\ifx\csname natexlab\endcsname\relax\def\natexlab#1{#1}\fi
\expandafter\ifx\csname url\endcsname\relax
  \def\url#1{\texttt{#1}}\fi
\expandafter\ifx\csname urlprefix\endcsname\relax\def\urlprefix{URL }\fi
\providecommand{\eprint}[2][]{\url{#2}}

\bibitem[{Flaugher(2012)}]{decam}
Flaugher, B. 2012, in Society of Photo-Optical Instrumentation Engineers (SPIE)
  Conference Series, vol. 8446 of Society of Photo-Optical Instrumentation
  Engineers (SPIE) Conference Series, 35

\bibitem[{Marshall \& DePoy(2005)}]{ff2005}
Marshall, J.~L., \& DePoy, D.~L. 2005, ArXiv Astrophysics e-prints.
  \eprint{arXiv:astro-ph/0510233v1}

\bibitem[{Rheault et~al.(2010)Rheault, DePoy, Behm, Kylberg, Cabral, Allen, \&
  Marshall}]{decal2010}
Rheault, J.-P., DePoy, D.~L., Behm, T.~W., Kylberg, E.~W., Cabral, K., Allen,
  R., \& Marshall, J.~L. 2010, in Society of Photo-Optical Instrumentation
  Engineers (SPIE) Conference Series, vol. 8446 of Society of Photo-Optical
  Instrumentation Engineers (SPIE) Conference Series, 201

\bibitem[{Rheault et~al.(2012)Rheault, DePoy, Marshall, Prochaska, Allen, Wise,
  Martin, \& Williams}]{decal2012}
Rheault, J.-P., DePoy, D.~L., Marshall, J.~L., Prochaska, T., Allen, R., Wise,
  J., Martin, E., \& Williams, P. 2012, in Society of Photo-Optical
  Instrumentation Engineers (SPIE) Conference Series, vol. 8446 of Society of
  Photo-Optical Instrumentation Engineers (SPIE) Conference Series, 252

\bibitem[{Stritzinger et~al.(2011)Stritzinger, Phillips, Boldt, Burns,
  Campillay, Contreras, Gonzalez, Folatelli, Morrell, Krzeminski, Roth,
  Salgado, DePoy, Hamuy, Freedman, Madore, Marshall, Persson, Rheault,
  Suntzeff, Villanueva, Li, \& Filippenko}]{stritz}
Stritzinger, M.~D., Phillips, M.~M., Boldt, L.~N., Burns, C., Campillay, A.,
  Contreras, C., Gonzalez, S., Folatelli, G., Morrell, N., Krzeminski, W.,
  Roth, M., Salgado, F., DePoy, D.~L., Hamuy, M., Freedman, W.~L., Madore,
  B.~F., Marshall, J.~L., Persson, S.~E., Rheault, J.-P., Suntzeff, N.~B.,
  Villanueva, S., Li, W., \& Filippenko, A.~V. 2011, AJ, 142, 156

\bibitem[{Stubbs et~al.(2010)Stubbs, Doherty, Cramer, Narayan, Brown, Lykke,
  Woodward, \& Tonry}]{panstarrs}
Stubbs, C.~W., Doherty, P., Cramer, C., Narayan, G., Brown, Y.~J., Lykke,
  K.~R., Woodward, J.~T., \& Tonry, J.~L. 2010, ApJS, 191, 376

\bibitem[{Stubbs et~al.(2007)Stubbs, Slater, Brown, Sherman, Smith, Tonry,
  Suntzeff, Saha, Masiero, \& Rodney}]{stubbs2007}
Stubbs, C.~W., Slater, S.~K., Brown, Y.~J., Sherman, D., Smith, R.~C., Tonry,
  J.~L., Suntzeff, N.~B., Saha, A., Masiero, J., \& Rodney, S. 2007, in The
  Future of Photometric, Spectrophotometric and Polarimetric Standardization,
  edited by C.~Sterken. (San Francisco: Astronomical Society of the Pacific),
  vol. 364 of ASP Conference Series, 373

\end{thebibliography}

\end{document}